\documentstyle[12pt]{article}
\begin{document}
{\hfill {LANDAU-95-GR-1}}

\vspace{0.5cm}

\centerline{\bf MODERN COSMOLOGICAL MODELS WITH DARK MATTER}
\centerline{\bf AND THEIR CONFRONTATION WITH OBSERVATIONAL DATA}
\bigskip

\centerline{A.A. Starobinsky}
\centerline{Landau Institute for theoretical Physics,}
\centerline{Russian Academy of Sciences, Moscow 117334, Russia}

\vspace{0.5cm}
\noindent
{\it Short version of the plenary talk
presented at the First International Conference on Cosmoparticle Physics
"Cosmion-94" (Moscow, 5-14 December 1994). Published in: Cosmoparticle
Physics. 1, eds. M.Yu. Khlopov, M.E. Prokhorov, A.A. Starobinsky, J. Tran
Thanh Van, Edition Frontiers, 1996, pp. 141-146.}

\vspace{0.5cm}
\noindent
{\bf ABSTRACT.} 
New systematic classification of cosmological models of
the present Universe is introduced. After making the comparison of these
models with all existing observational data three viable models remain:
the cold dark matter model with the cosmological constant (which becomes
the most reasonable one if the Hubble constant $H_0>60$ km/s/Mpc); the mixed
cold-hot dark matter model (models with two and especially three types of
neutrinos with equal masses are in a slightly better agreement with
observational data than the model with one massive neutrino); the pure
cold dark matter model with a step-like initial spectrum of perturbations.
The two latter models require $H_0\leq 60$ km/s/Mpc.

\vspace{0.5cm}

Looking at the enormous amount of papers in which different cosmological
models of the present Universe are studied and compared with observational
data, the first impression might be that there exists a great degree of
freedom and arbitrariness in such models. However, actually this is not so.
First, using a natural fundamental classification described below, it
appears that all currently popular models follow some simple logic. It becomes
clear also what are the natural directions of generalization of these models.
Second, if {\em all} existing observational data are used (and not some
arbitrarily chosen tests only), then a number of viable cosmological models
reduces to a few - we shall argue below that only three models remain viable at
present. Would we know the Hubble constant $H_0$ with a better accuracy, we
could further reduce this number and might even finish with a unique preferred
one.

To introduce this classification, let us remind the logical structure of
cosmology. The aim of theoretical cosmology, as of any other branch of science,
is to produce definite predictions basing upon as minimum number of assumptions
as possible ("Occam's razor") which can be tested by observations. To make
these predictions, first, we need equations (or Lagrangians). It is generally
accepted now that the Universe is described by a space-time metric
satisfying the Einstein equations with some matter source. Moreover, it
was shown that a broad class of theories more general than the Einstein
gravity (e.g., scalar-tensor theories of gravity, the $R+f(R)$ theory
where $R$ is the trace of the Ricci tensor, etc.) can be represented in the
Einsteinian form after some conformal transformation. So, the left-hand side
of the equations may be taken as $R_{ik}-{1\over 2}Rg_{ik}$.

However, the equations are not defined if their right-hand side is not
completely specified. So, we should assume some matter content of the Universe
(in particular, at present time). It is remarkable that all viable cosmological
models require the main part of matter in the present Universe to be the
dark non-baryonic matter. It is dark in the direct sence of the word,
i.e. not participating in electromagnetic interactions. Moreover, all
candidates for this dark matter are not yet discovered in laboratory
experiments (or not yet discovered are the specific properties of particles
making them important for cosmology, if speaking, e.g., about neutrino
masses). Thus, modern cosmology strongly suggests and awaits great
laboratory discoveries in the field of particle physics!

The evidence for dark matter follows from the following numbers. \\
1. The energy density of luminous matter in the Universe (stars, hot gas
in galaxies and between them) in terms of the critical one
$\varepsilon_c=3H_0^2/8\pi G$ (we assume $c=\hbar=1$): $\Omega_{lum}< 0.01$. \\
2. The primordial (Big Bang) nucleosynthesis (BBN) prediction for
the energy density of baryons in the Universe based on observed
abundances of light elements ($^4$He, D, $^3$He, $^7$Li) is
$\Omega_b=(0.01-0.02)h^{-2}$ where $h=H_0/100$.   \\
3. The virial energy density of matter gravitationally clustered at scales
$R=(5-25)h^{-1}$ Mpc is $\Omega_{vir}=0.2-0.5$.  \\
4. Finally, the theoretical expectation in most inflationary models
is $\Omega_{tot}=1$. \\
Comparing 2 with 1, we see that some amount of baryon dark matter
should exist (e.g., in the form of faint stars). However, from the
comparison of 3 to 1 we see that much more matter should be in a
non-baryonic dark form. Note that this conclusion is not based on any
theoretical assumptions about a cosmological model. If we compare 4
with 3 (using now some hypothesis about the early Universe), then
the further conclusion follows that there may exist different types of
dark non-baryonic matter in the Universe.

The main 3 types of dark matter (further, we shall speak about non-baryonic
dark matter only) are: \\
1. Hot dark matter (HDM) - neutrinos with restmasses of the order of few eV. \\
2. Cold dark matter (CDM) - sypersymmetric particles with masses $\sim
100$ Gev or axions with $m_a\sim 10^{-5}$ eV. \\
3. Ultra-cold (vacuum-like) matter - the cosmological constant.  \\
However, we understand now that there are no impenetrable barriers
between these 3 types. In more exotic models, intermediate types of dark
matter may appear between these ones. In particular, warm dark matter
(e.g., thermal particles with masses of about $1$ keV or non-equilibrium
neutrinos) fills the gap between 1 and 2, and a scalar field with the
exponential potential can be used to produce an arbitrary equation of state
of matter between 2 and 3 (actually, even in the range from $p=\varepsilon$
to $p=-\varepsilon$ where $p$ is the pressure of matter).

So, if the dark matter content is specified, the equations are fully
determined. However, to solve them we need initial conditions (mostly,
for perturbations). The observational evidence from $\Delta T/T$
angular anisotropy of the cosmic microwave background (CMB) proves that the
Universe was FRW-like (isotropic and homogeneous) at the recombination
time (at the redshift $z\approx 1100$). The BBN theory shows that the
Universe was FRW-like in the much earlier period ($t=(1-100)$ s after the
cosmological singularity, $z=10^8-10^9$). The absence of primordial
black holes with $M=10^{15}-10^{17}$ g which would evaporate now through
the Hawking radiation suggests the isotropy of the Universe at even
earlier times, and so on. However, even if we assume that the Universe
was FRW-like just from the very beginning, i.e. from times close to the
Planck time $t_P=\sqrt G$, we still remain with three arbitrary functions
of space coordinates (or 3 arbitrary functions of the wave vector ${\bf k}$
in the Fourier representation) specifying initial amplitudes of the
quasi-isotropic modes (called so because they do not destroy local isotropy
and homogeneity at arbitrarily early times). One of these functions refers
to scalar (adiabatic) perturbations and two others - to the quasi-isotropic
mode of tensor perturbations (primordial gravitational waves). In more
complicated models, additional functions giving initial values for
non-decreasing isocurvature modes can appear. Therefore, generally we have a
large functional arbitrariness in initial conditions.

The second main advantage of the inflationary scenario of the early Universe,
after the elegance and beauty of its main assumption that our Universe
was in the maximally symmetric (de Sitter, or inflationary) state during some
period in the past, is that it predicts these initial conditions in terms of a
small number of fundamental parameters of an effective Lagrangian describing
an inflationary stage realized in any concrete version of this scenario. Thus,
the arbitrariness reduces to a few (minimum one) parameters. From all this
discussion, a natural and fundamental classification of cosmological models
of the present Universe follows (including non-inflationary models, too, but I
shall discuss only inflationary ones further)[1] (see also [2]): \\
let us classify them by their level of complexity number which is equal to a
number of  \\
a) new (not known before, e.g., $H_0$ and $\Omega_{\gamma}$ are
not counted), \\
b) fundamental (appearing in basic equations, not in initial
conditions), \\
c) significant (more than $\sim 10\%$ effect, e.g. if
$|n_s-1|\le 0.1$, it is counted as $n_s=1$), \\
d) dimensionless (this can be always achieved, e.g., working in the Planck
units) \\
{\em constants} introduced in any particular models in order to explain all
observational data. These constants may refer {\em either} to the present dark
matter content {\em or} to the initial spectrum of perturbations. Note that the
complexity level of a given model may decrease as a result of future
discoveries
in laboratories (e.g., measurements of neutrino restmasses) or due to a
progress in an underlying unified physical theory (e.g., a derivation of
inflaton parameters from the superstring theory).

Distinguishing features of this classification which
radically discriminate it from all previous attempts in this direction
are, first, the fact that the number of assumptions using no numbers
(numerical constants) is not counted at all. Thus, this classification
is favourable even for very "strange", "crazy" models if they are internally
consistent and do not introduce a large number of free parameters. Second,
it counts assumptions referring to the present dark matter content and to the
initial spectrum of perturbations on equal footing. Thus, if we want to
complicate a model under the pressure of observational data and shift to
the next level of complexity, we may add one parameter eigher to the
effective Lagrangian describing an inflationary stage (and then we get one
more parameter in the initial spectrum of perturbations), or to the
description of the dark matter content (e.g., by introducing one new type of
dark matter). Then the logical order of the development is to begin
with the lowest complexity level, compare it with the data, and if there
is no agreement, move to successive higher levels until the agreement
will be reached.

In the Table 1 below a very brief sketch of existing cosmological tests
is presented with an approximate range of scales to which they are
sensitive (or of scales for which data exist at present).  Here $\Phi$
is the gravitational potential and $h_{\alpha \beta}$ are the tensor
perturbations (gravitational waves). The Table 2 gives a list of models
with their complexity level numbers, initial (fundamental) parametres,
observational parameters which can be unambiguously expressed through the
fundamental ones, and the result of comparison with observational data.
$A$ is the {\it rms} amplitude of the spectrum of adiabatic perturbations,
$A_{iso}$ is the same for isocurvature perturbations, $\sigma_8$ is the
total rms top-hat matter perturbation at the scale $R=8h^{-1}$ Mpc. For
the model 6, $\phi_0$ is the value of the inflaton field at the moment of
bubble formation, $\phi_f$ is its value at the end of inflation. For the
model 10, $\phi_s$ is the value of the inflaton field at the beginning of
the last stage of inflation. For the model 11, $\nu1$ denotes an unstable
neutrino while $\nu$ denotes the stable one. The latter model (which is
actually a class of different models) is still largely unexplored, so no
conclusion about its viability is drawn.

All details about the confrontation of the models with observational
data can be found in [3,2] for the models 3 and 9, in [4,5] for the
model 4, in [6] for the model 5 and in [7,8] for the model 10. The
list of references is, of course, very incomplete, but references to
other papers can be found in the given ones.

So, the first remarkable conclusion is that it is possible to explain
all existing cosmological observational data using 2 or 3 parameters
only (4 in the case of three neutrinos with comparable masses). There no
necessity to go to higher complexity levels at present.

The second conclusion shows the dependence of results on the value of $H_0$
(still not determined with the desired degree of accuracy). If $H_0>60$
km/s/Mpc, then the best (and probably the only possible) model is the
model 5: CDM$+\Lambda$ with $n_s\approx 1$. So, a reliable observational
proof that $H_0>60$ should be considered as a very strong argument for the
positive cosmological constant. Note, however, another prediction that
$H_0<80$ because in the opposite case no reasonable model exists (at least, at
the complexity levels considered). On the other hand, for $H_0\le 60$
we can avoid introducing the cosmological constant, and then the choice
is between the mixed CDM+HDM models 3 or 4 and the pure CDM model 10.
They can be most easily discriminated by the abundance of compact objects at
large redshifts.

The third conclusion is that if the mixed CDM+HDM model is the right one,
then it is very interesting that the model 4 having several neutrino species
with comparable masses produces a fit to all data slightly better than the
model 3 with one massive meutrino. So, in this case cosmology provides
some support (though not the final evidence, of course) for strong neutrino
mixing.

This research was partially supported by the Russian research project
"Cosmomicrophysics" through Cosmion and by the INTAS grant 93-3364.

\vspace{0.5cm}

\centerline{\bf REFERENCES}

\vspace{0.5cm}

\noindent
1. A.A. Starobinsky. In: {\it Theoretical and Experimental Problems
of Gravitation} (Abstr. 8th Russian Grav. Conf., Pushchino, 25-28 May
1993), Moscow, 1993, p. 149. \\
2. D.Yu. Pogosyan, A.A. Starobinsky. Astroph. J., {\bf 447}, 465, 1995. \\
3. D.Yu. Pogosyan, A.A. Starobinsky. MNRAS, {\bf 265}, 507, 1993.  \\
4. J. Primack, J. Holtzman, A. Klypin and D. Caldwell. Preprint
astro-ph/9411020, 1994; Phys. Rev. Lett., in press, 1995.  \\
5. D.Yu. Pogosyan, A.A. Starobinsky. In: {Large Scale Structure in the
Universe} (Proc. of the XI Potsdam Workshop on Relativistic Astrophysics,
Potsdam, 19-24 Sept. 1994), eds. J.P. M\"ucket et al., World Scientific,
Singapore, 1995; preprint astro-ph/9502019, 1995. \\
6. L.A. Kofman, N.Yu. Gnedin, N.A. Bahcall. Astroph. J., {\bf 413}, 1, 1993. \\
7. S. Gottl\"ober, J.P. M\"ucket, A.A. Starobinsky. Astroph. J., {\bf 434},
417, 1994. \\
8. P. Peter, D. Polarski, A.A. Starobinsky. Phys. Rev. D, {\bf 50}, 4827,
1994.

\pagebreak
\centerline{Table 1. Tests}

\bigskip

\begin{tabular}{|c|c|c|}
\hline
Name & Type of perturbation &  Scales ($h^{-1}$ Mpc)\\
\hline
Large-angle $\Delta T/T$ ($\theta>2^0$)& $\Phi , h_{\alpha\beta}$& 200-6000\\
Intermediate angle $\Delta T/T$ ($10'<\theta<2^0$)& $\Phi, \nabla \Phi,\Delta\Phi$& 20-200\\
Peculiar (bulk) velocities&$\nabla\Phi$&10-75\\
Galaxy-galaxy correlations&$ \Delta\Phi$&0.1-200\\
Large-scale structure&${\partial^2\Phi\over\partial x^{\alpha}\partial x^{\beta}}$&5-100\\
Cluster-cluster correlations&$ \Delta\Phi$ &5-150 \\
Cluster abundance &$ \Delta\Phi$ & 5-10 \\
Galaxy and quasar abundance at large z &$\Delta\Phi$& 0.5-1 \\
$Ly-\alpha$ clouds& $\Delta\Phi$& 0.1-1\\ \hline
\end{tabular}

\pagebreak

\centerline{Table 2. Classification of cosmological models}

\bigskip

\begin{tabular}{|c|c|c|c|c|c|}
\hline
No.&Models & Level No.& Fundamental&Observational&Agreement with\\
   &      &          &parameters   &parameters  &observational\\
   &      &          &               &             &data\\
\hline
1 & Standard CDM & 1 &$(H^2/\dot\phi)_{inf}$& A & No \\
  & ($n_s=1$)    &   &                      &   &    \\
2 & CDM, $n_s=-3$ & 1 &  & $A_{iso}$& No \\
  & isocurvature & & & & \\
3 & CDM+HDM, & 2 & $(H^2/\dot\phi)_{inf}$, & $A, \Omega_{\nu}$& Marginally \\
 &$1\nu$, $n_s=1$, &   & $m_\nu/M_p$             &               & good if\\
 &standard      &   &                         &             &$H_0\simeq50$,\\
 & concentration &   &                         &      &$\Omega_\nu\simeq0.2$\\
4 &CDM+HDM &3,4&$(H^2/\dot\phi)_{inf}$, & $A, \Omega_{\nu_i}$& Good if\\
 & $2\nu$ or $3\nu$, &  & $m_{\nu_1}=m_{\nu_2}$ &       & $H_0 \leq 60$,\\
 &$n_s=1$,& & $(=m_{\nu_3})$ & &$\sum m_{\nu_i}\simeq5(H_0/50)^3$ eV\\
 & standard &     &                 &             &   \\
 & concentration &  &               &             &   \\
 5 & CDM+$\Lambda$, & 2 & $(H^2/\dot\phi)_{inf},$ &$A,\Omega_\Lambda$ & Good,\\
 &$\Omega_{tot}=1,$ &  & $\Lambda M^2_p$ &         & $H_0=50-80$,\\
 & $n_s=1$ &  &                        &         &$\Omega_m=0.5-0.2$\\
 6 & CDM curved, & 2 & $(H^2/\dot\phi)_{inf}$,
 &$A,\Omega$&Worse than\\ &$\Omega_m<1$, &  &$\phi_0/\phi_f$ &  &
 CDM+$\Lambda$,\\
 &$n_s=1$      &   &                &  & $\Omega_m>0.3$, \\
 &           &  &                 &  & $H_0<70$\\
 7 &CDM & 2 &$V(\phi)=V_0e^{-\alpha{\phi\over M_p}}$, & $A(k_{hor})$, &No\\
 & tilted  &  & $\alpha,V_0$         & $n_s\neq1$ &(if $\sigma_8\le 0.7$) \\
 & adiabatic &  &    &  & \\
 8 &CDM & 2 &  &$A_{iso}(k_{hor})$, & No \\
 & tilted &  &  & $n_s\neq-3$ &  \\ & isocurvature &  &  &  & \\
 9 &CDM+HDM &3& $\alpha,V_0$, & $A(h_{hor})$, & No, if \\
 & tilted, $1\nu$,&  & $m_\nu/M_p$ & $n_s\neq1,$        & $|n_s-1|>0.1$ \\
 & standard &  &   & $\Omega_\nu$ &    \\
 & concentration&  &            &              &  \\
 10 &CDM with & 3 &$(H^2/\dot\phi)_{+}$, &$A_{+}, A_{-}$,& Good if \\
 & a step-like &  & $(H^2/\dot\phi)_{-}$, & $ k_s$ &  $H_0 \leq 60$ \\
 & spectrum & & $\phi_s/\phi_f$ &      &   \\
 11 & CDM+HDM, & 3 & $(H^2/\dot\phi)_{inf}$,& $A,\Omega_\nu$,& \\
 &decaying $\nu$,&  & $ m_{\nu_1},\tau_{\nu_1}$ & $ m_\nu$ &  \\
 & $n_s=1$ & &    &    &   \\ \hline
\end{tabular}
\end{document}